# Floating-zone growth and property characterizations of high-quality La$_{2-x}$Sr$_x$CuO$_4$ single crystals

Xiaoli Shen, Zhengcai Li, Caixia Shen, Wei Lu, Xiaoli Dong, Fang Zhou[*],
Zhongxian Zhao

National Laboratory for Superconductivity, Institute of Physics and Beijing National Laboratory for Condensed Matter Physics, Chinese Academy of Sciences，Beijing 100190, P. R. China

[*] Corresponding author: E-mail fzhou@ssc.iphy.ac.cn; Phone (86)-10-8264 9526; Fax (86)-10-8264 9486

**Abstract**

We have grown underdoped (x = 0.11, 0.12) and optimally doped (x = 0.16) La$_{2-x}$Sr$_x$CuO$_4$ (LSCO) crystals by traveling-solvent floating-zone (TSFZ) technique. In order to obtain LSCO single crystals of high quality, we have made much effort to optimize the preparation procedures. For example, we have adopted sol-gel route to prepare highly fine and homogeneous LSCO precursor powder, and used quite slow growth rate. The sizable grown crystal ingots are typically about 6.6 mm in diameter and 100 mm in length, and no impurity phases were detected. The high quality of grown crystals has been verified by double-crystal x-ray rocking curves, with full-width-at-half-maximum (FWHM) being only 113 ~ 150 arcseconds (or 0.03° ~ 0.04°), which are the best data reported so far for LSCO crystals. The superconducting critical temperatures (T$_C$'s) of the cuprate crystals are 30 K for x = 0.11, 31 K for x = 0.12 and 38.5 K for x = 0.16 samples according to magnetic measurements.

1. Introduction

Among the cuprate superconductor family, the $La_{2-x}Sr_xCuO_4$ (LSCO) system is known as the one of fewer components and of a simpler layered structure of $K_2NiF_4$ type with single $CuO_2$ plane. Its overall density of doped charge (holes) can be continuously tuned by substitution of $Sr^{2+}$ cations for $La^{3+}$ ones, and its superconductivity, which evolves from insulating antiferromagnetic region with strontium doping, occurs in a wide doping range of x = 0.06 ~ 0.26 [1]. Because of these advantages, LSCO is considered as a good candidate for studying the origin of high-$T_C$ superconductivity in strongly correlated electron system since the discovery of cuprate superconductor. Moreover, some interesting results have recently been reported for LSCO system. For example, a composite charge model has been proposed based on far-IR experiments that only a small fraction of doped holes take part in the free charge transport while the others form a 2D electronic lattice [2]. It has also been found that there exist only two distinct intrinsic superconducting phases with $T_C$'s of either 15 K or 30 K at so-called 'magic' doping levels of x = 1/16 (0.0625) and 1/9 (~0.11), respectively, in the underdoping regime of LSCO [3-5]. And these two intrinsic $T_C$'s are found to be quite field independent up to 5 Tesla [5]. These facts imply that LSCO may serve as an important prototype for revealing the mechanism of high-$T_C$ superconductivity.

Single crystal sample of high quality can exhibit truly intrinsic property and anisotropy of the system, it is therefore indispensable for the fundamental research. Traveling-solvent floating-zone (TSFZ) method is now accepted as a unique technique to grow the incongruent melting cuprate crystals. We report in this paper the results of TSFZ growth of LSCO crystals and characterization of their properties. In order to prepare high-quality LSCO crystals, we have taken various measures to optimize the whole preparation process. Those include: (1) sol-gel technique is adopted to synthesize very fine and homogeneous LSCO precursor powder with its final heating temperature no higher than 850 °C. By using this sol-gel precursor powder, LSCO ceramic feed rod can be sintered very densely, and this is of great benefit to a perfect solid-liquid interface and a stable molten zone during TSFZ growth; (2) a slow growth rate of 0.6 mm/h is used and the whole growth process is manipulated very carefully; (3) the grown crystals are checked by various measurements to insure the improvement in crystal quality. For example, the FWHM of x-ray rocking curve is as good as 0.03° ~ 0.04° for LSCO crystals grown in this work, that is the best data reported so far for LSCO crystals [3,6].

## 2. Experimental

High purity (better than 99.99%) oxides $La_2O_3$, $CuO$ and carbonate $SrCO_3$ were used as raw materials based on the chemical formula of $La_{2-x}Sr_xCuO_4$, but with $CuO$ being in excess by ~2 mol% to compensate its evaporation loss during the growth process. The $La_2O_3$ oxide was heated at 900 ~ 1000 °C for 6 ~ 10 hours and $CuO/SrCO_3$ were heated at 125 °C overnight just before weighing. In this work, a sol-gel method was adopted to prepare very fine and homogeneous LSCO precursor powder through following steps:

(i) Proper quantity of distilled water was first put into a 1000 ml beaker containing the mixture of the raw materials (totally 110 ~ 125 g for each batch), then nitric acid (concentration 65%) was added slowly into the beaker with stirring. This step should be done with caution to avoid spattering of the solution due to decomposition of $SrCO_3$. The quantity of the nitric acid used was calculated according to the mole numbers and valences of all the metal elements, usually with an excess of about 10 ml. The solution was heated at 80 °C for about 10 hours until all the chemicals were dissolved completely. A blue, transparent nitrate solution was obtained.

(ii) Citric acid (chelating agent) and Ethylenediamine were used to form metal-citrate complex sol. A proper amount of citric acid was weighed based on the mole numbers of the metals. The powder of citric acid was first dissolved in distilled water, then the citric acid solution was added into the nitrate solution obtained in (i) with stirring. When ethylenediamine was dropped into the solution with stirring by machine, the blue, transparent solution first became light blue in colour and full of floccules, it then gradually changed into a purple, clean and uniform sol solution without any floccules. Thus obtained sol was checked by a pH meter to be slightly alkaline (pH ~ 7.6) and was ~ 700 ml in volume.

(iii) A thick polymerized complex gel was obtained after the sol was heated at 60 ~ 80 ℃ for about 10 hours in a big alumina crucible with stirring by machine. After heated again at 140 ℃ for 3 hours in a muffle furnace to further remove the water, the gel changed into dried porous blocks. Then the blocks were crushed into small pieces and heated up to 350 ℃ in a ventilated muffle furnace to remove most organic species. This heating procedure took about 120 hours, including temperature increasing and dwelling (final dwelling being at 350 °C for 15 hours), and the temperature should be increased very slowly to prevent the material from burning. This step is important for controlling the content of carbon residual in the material to a level less than 1/1000, as checked by high-frequency combustion/infrared measurement. A light brown fine powder was obtained after ball milling of 2 hours.

(iv) The light brown powder was then heated at temperatures of 600 °C and 850 °C for 15 hours each. Again the temperature should be raised slowly. After each heating, the powder was ground with a laboratory planetary ball mill for 6 ~ 8 hours. Then a black precursor powder of pure LSCO phase was synthesized, as checked by XRD analysis.

The LSCO precursor powder was put into a silicone tube and pressed with a home-made hydrostatical press under a pressure > 600 MPa to form a very compact cylindrical rod. The rod had a dimension of about 8.5 mm in diameter and 135 mm in length. Then it was suspended with nickel wire in a specially designed vertical furnace (Crystal System Corp., VF-1800/EF-6000) and was sintered at 1240 ~ 1280 °C for 2 hours depending on different compositions. During sintering the rod was kept rotating and moving up and down to have an identical thermal history for the whole rod. Thus prepared LSCO ceramic rod was very dense and uniform, mainly due to the use of sol-gel precursor powder which was very fine and homogeneous and was synthesized at temperature no higher than 850 °C.

The solvent was prepared by traditional solid state reaction. The composition was of 80 ~ 84 mol% CuO as self-flux. We took into account of the previously reported result of distribution coefficients $k_{Sr}$ of Sr doping into $La_2CuO_4$ [7]. The mixture of chemical powders was heated twice at 980 ℃ for 15 ~ 20 hours followed by groundings before pressed into small pellets. The pellets were then sintered at 1000 ~ 1020 °C for 15 ~ 20 hours depending on each composition. In order to reduce bubbles that may emerge during the growth process, the solvent pellets should be sintered at high enough temperature but at which be sure that they are not molten.

An infrared-heating floating-zone furnace with a quartet ellipsoidal mirror (Crystal systems Corp., Model FZ-T-10000-H) was used to grow the LSCO crystal by TSFZ method. LSCO crystals oriented along tetragonal [100] or [110] directions were used as seed crystals. The crystal growth was carried out under an oxygen pressure of 0.2 ~ 0.3 MPa to reduce the evaporation of Cu and at a quite slow zone travelling rate of 0.6 mm/h.

Powder X-ray diffraction (XRD) analysis was made on an MXP18A-HF diffractometer using Cu-Kα radiation, with 2θ ranging from $10°$ to $80°$ and a 2θ scanning step of $0.01°$. The crystal compositions were checked by the inductively coupled plasma atomic emission spectroscopy (ICP-AES). The experiments of double-crystal x-ray rocking curves were preformed on a diffractometer (Bede, model D1) with Cu $K_{\alpha 1}$ radiation reflected from Si (220) monochromator. The magnetic susceptibility measurements were carried out on a Quantum Design MPMS XL-1 system.

### 3. Results and discussion

Figure 1a is a photo taken during the TSFZ growth of LSCO crystal, showing a very stable molten zone, perfect solid-liquid interfaces as well as little penetration of liquid into the ceramic feed rod. These favourable conditions, being as a result of the use of highly dense and uniform ceramic feed rod and careful operations, are very helpful for growing high-quality crystals. The size of grown crystal ingots is typically about 6.6 mm in diameter and 100 mm in length, as shown in figure 1b.

Figure 2 is an example of back-scattering x-ray (tungsten target) Laue patterns used for crystal orientation, which is a result obtained on (001) plane (ab plane) of LSCO x = 0.11 crystal and shows very sharp and uniform diffraction spots with the 4-fold axis symmetry. The crystalline perfection of the LSCO crystals (with doping x = 0.11, 0.12, 0.16) grown in this work was checked by double-crystal x-ray rocking curves of (008) Bragg reflection, as plotted in Figure 3. The full-width-at-half-maximum (FWHM) was as small as 113 ~ 150 arcseconds (or $0.03°$ ~ $0.04°$), indicating that the LSCO crystals are of high crystalline quality. As far as we know, this is the best FWHM data reported for LSCO crystals.

In order to check the phase purity and strontium doping uniformity along the whole grown ingots, we did careful XRD analysis and composition measurements by ICP-AES as follows. On each grown ingot with doping x = 0.11, 0.12 and 0.16, sectioned from different positions were two slices of thin disc-shaped crystal of 1mm thickness, which were designated as sample A and sample B. Sample A was located near the backend of the ingot where the crystal growth finished, and sample B was away from sample A at an interval of 6 ~ 8 cm. Sample A and B were ground into fine powders before they were measured by XRD and then by ICP-AES. Given in figure 4 are the XRD patterns for the three pairs of samples with x = 0.11, 0.12 and 0.16. All the samples are of pure LSCO phase (all the diffraction peaks can be indexed with appropriate tetragonal unit cell) and no impurity phases are detected within the experimental resolution. Listed in table 1 are the results of ICP-AES measurements that the compositions in grown crystals are the same as that in the ceramic feed rods as expected and no variations in Sr doping are observed for each pairs of the samples. These results show that the grown ingots are uniform in composition and phase purity.

The Meissner (field-cooled) and the shielding (zero-field-cooled) signals of the crystals are shown in figure 5. The data were corrected for demagnetizing factors [8]. All the samples were processed into a regular dimension of 2 mm×1 mm×0.3 mm and were measured with c axis along the field. The shielding volume fractions of all the samples are close to 100%, indicating that there are no macroscopic inhomogeneity and/or weak links. The onset superconducting transition temperatures ($T_C$'s) were determined from the differential curves of the Meissner signals (the same way as described in reference 5), with the $T_C$'s being 30 K, 31 K and 38.5 K for x = 0.11, 0.12 and 0.16 crystals, respectively. Moreover, both sample A and sample B of x = 0.11 crystal were checked by the magnetic measurements. The two sets of results were of the same $T_C$ values as expected, again indicating the crystalline uniformity along the ingot.

## 4. Conclusion

High quality and sizable $La_{2-x}Sr_xCuO_4$ crystals with x = 0.11, 0.12 and 0.16 were grown by traveling-solvent floating-zone technique. The crystal quality and properties of the grown ingots were characterized by various methods such as x-ray diffractions (powder XRD, Laue patterns, rocking curves), composition analysis of ICP-AES and magnetic measurements. The high crystal quality was evidenced by double-crystal x-ray rocking curves with FWHM being only 113 ~ 150 arcseconds ($0.03°$ ~ $0.04°$), which are the best data reported so far for LSCO crystals. The grown crystal ingots were uniform in composition and phase purity. The onset superconducting transition temperatures are 30 K, 31 K and 38.5 K for x = 0.11, 0.12 and 0.16 crystals, respectively.

The use of highly fine and homogeneous LSCO precursor powders synthesized by sol-gel route as well as other optimized preparation conditions turned out to be quite effective in producing very dense and uniform ceramic feed rods and high-quality crystals. A dense and uniform feed rod is of great benefit to a perfect solid-liquid interface (ex. little penetration of liquid into the feed rod) and a stable molten zone during TSFZ growth, that also makes it possible to use a slow growth rate. All those are important for growing high-quality crystals by TSFZ method.

## Acknowledgements

We are grateful to Prof. H. Chen and Mr. J. Zhang for their kind helps in the measurements of double-crystal x-ray rocking curves. We also thank Mrs. H. Chen for her help in powder XRD analysis, and Mr. H. Shi and Mr. L. Wang for the composition determinations by ICP-AES. This work was financially supported by the projects from the Ministry of Science and Technology of China, National Natural Science Foundation of China and Chinese Academy of Sciences.

**Figure captions**

Figure 1. (a) A photo taken during the TSFZ growth of LSCO crystal with x = 0.12. In the middle is the molten zone, the lower part is the grown ingot，and the upper part is the ceramic feed rod. (b) As-grown ingot of x= 0.11 crystal. The lower end is the seed crystal.

Figure 2. The x-ray (W target) Laue pattern of the (001) plane (the ab-plane) for x = 0.11 LSCO crystal. The four small black dots near the middle were used for locating the center of the Laue camera.

Figure 3. Double-crystal x-ray rocking curves of (008) Bragg reflection for all the three LSCO crystals (x = 0.11, 0.12 and 0.16), with FWHM being only 113 ~ 150 arcseconds ($0.03°$ ~ $0.04°$).

Figure 4. Powder x-ray diffraction patterns for the three pairs of sample A and B (see text) of the LSCO crystals with x = 0.11, 0.12, 0.16. All the patterns of sample A and B can be indexed on appropriate tetregonal unit cells: a ~ 3.780 Å and c ~13.21 Å for x = 0.11; a ~ 3.777 Å and c ~ 13.21 Å for x = 0.12; a ~ 3.774 Å and c ~ 13.22 Å for x = 0.16.

Figure 5. Meissner curves (open symbol lines) and the shielding curves (solid symbol lines) measured under a field of 5 Oe for LSCO crystals with different doping levels.

Table 1. The comparison of the compositions determined by ICP-AES for the three pairs of sample A and B (see text) of LSCO crystals with x = 0.11, 0.12, 0.16.

**Figure 1**

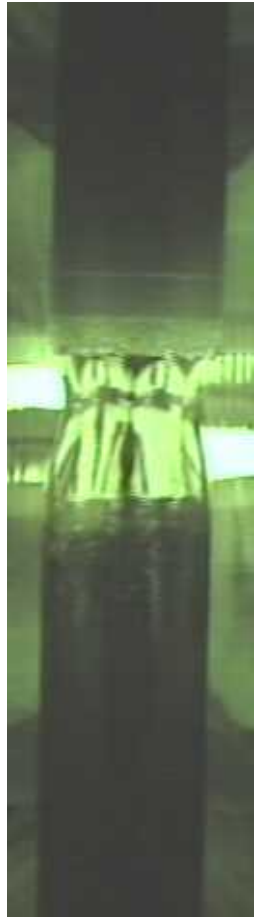 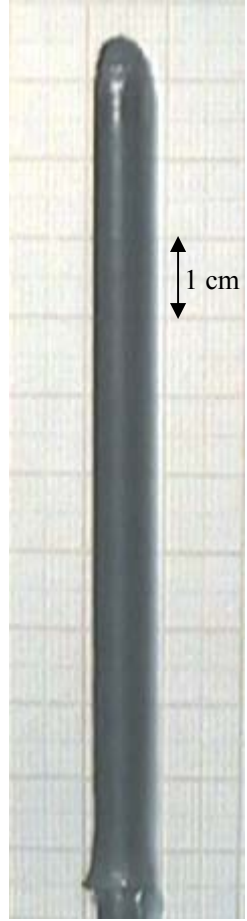

      **(a)**           **(b)**

**Figure 2**

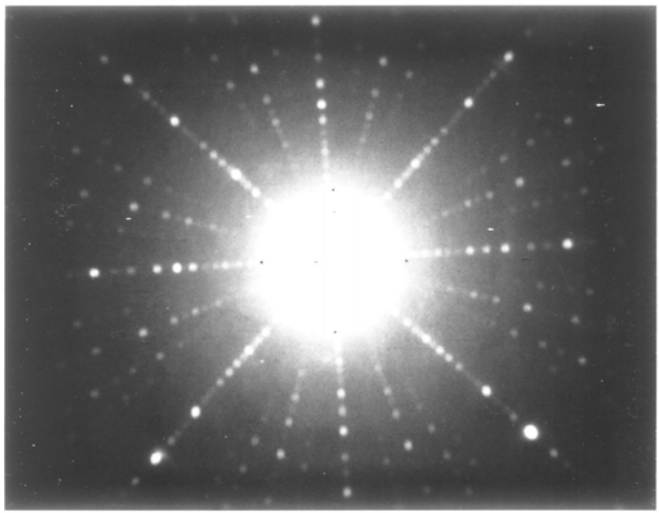

**Figure 3**

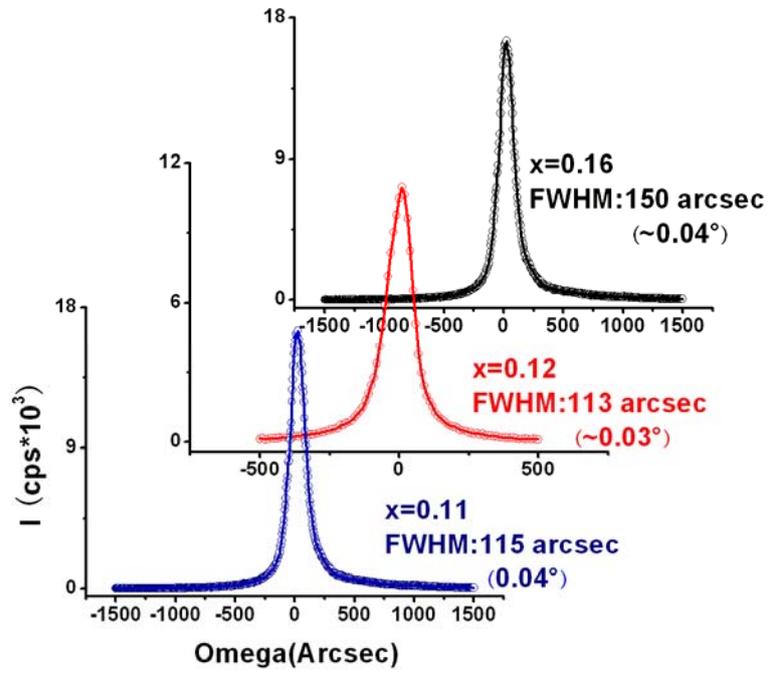

**Figure 4**

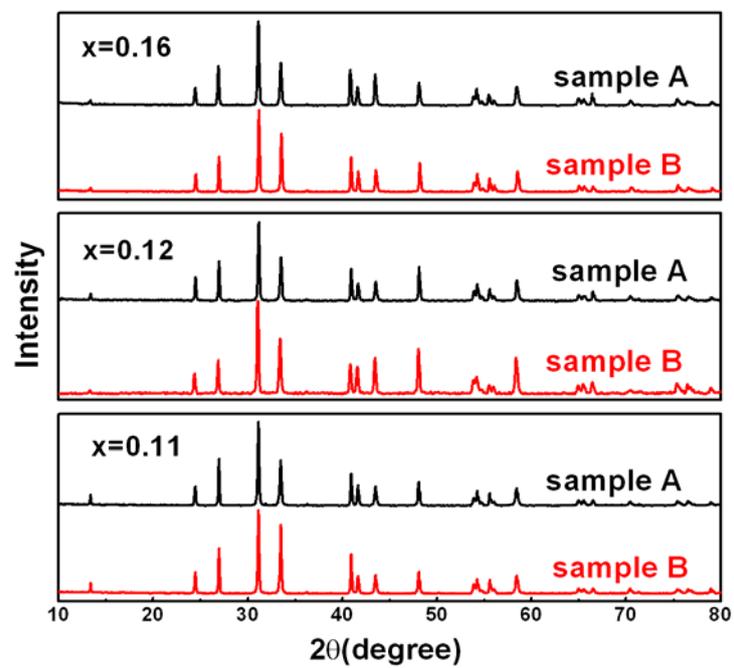

**Figure 5**

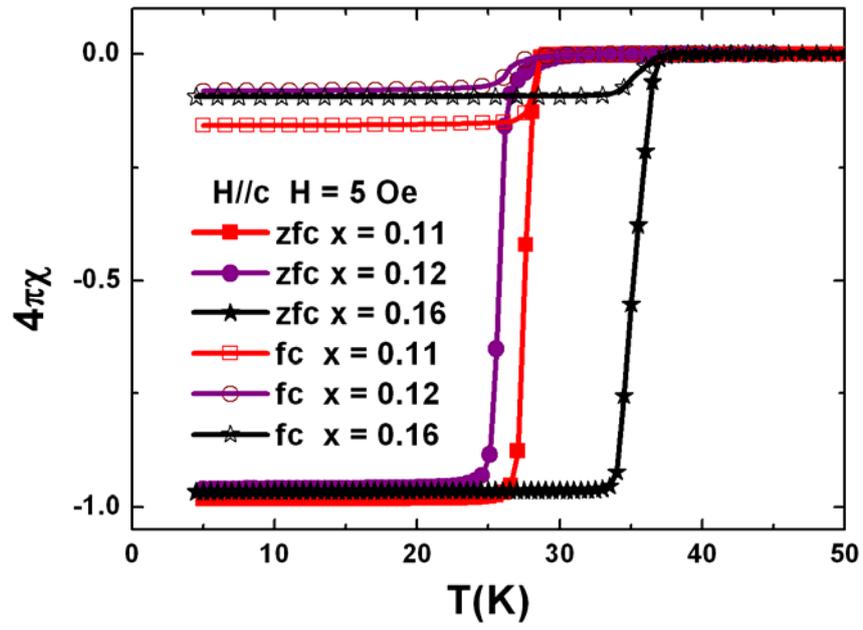

**Table 1**

| doping level<br>molar ratio | x = 0.11 | x = 0.12 | x = 0.16 |
|---|---|---|---|
| Cu:Sr (sample A) | 1:0.11 | 1:0.12 | 1:0.16 |
| Cu:Sr (sample B) | 1:0.11 | 1:0.12 | 1:0.16 |